\documentclass[12pt]{iopart}

\usepackage{iopams}
\usepackage{graphicx} 
\usepackage{color}
\usepackage{bm}

\begin{document}

\newcommand \be {\begin{equation}}
\newcommand \ee {\end{equation}}
\newcommand \bea {\begin{eqnarray}}
\newcommand \eea {\end{eqnarray}}
\newcommand \ve {\varepsilon}

\title[Phase transition in a dynamical decision model]{Symmetry-breaking phase transition in a dynamical decision model}

\author{Gaultier Lambert$^1$, Guillaume Chevereau$^{1,2}$ and Eric Bertin$^{1,2}$}

\address{$^1$ Universit\'e de Lyon, Laboratoire de Physique,
\'Ecole Normale Sup\'erieure de Lyon, CNRS,
46 all\'ee d'Italie, F-69007 Lyon, France\\
$^2$ IXXI Complex Systems Institute, 9 rue du Vercors, F-69007 Lyon, France
}

\ead{eric.bertin@ens-lyon.fr}

\begin{abstract}
We consider a simple decision model in which a set of agents randomly choose one of two
competing shops selling the same perishable products (typically food).
The satisfaction of agents with respect to a given store is related
to the freshness of the previously bought products.
Agents select with a higher probability the store they are most satisfied with.
Studying the model from a statistical physics perspective, both through numerical simulations and mean-field analytical
methods, we find a rich behaviour with continuous and discontinuous phase
transitions between a symmetric phase where both stores maintain the same
level of activity, and a phase with broken symmetry where one of the two shops
attracts more customers than the other.
\end{abstract}

\section{Introduction}

Phase transitions and symmetry breaking are standard phenomena
occuring in physical systems due to the presence of interactions between
the involved microscopic entities (most often atoms or molecules) \cite{Reichl}.
A prominent example is the
ferromagnetic Ising model \cite{Huang}, in which microscopic magnetic moments,
schematically represented as spin variables $s_i=\pm 1$, get aligned
at low temperature due to the interaction energy that couples neighbouring
spins.
Outside physics, phase transitions are also often found in complex systems 
composed of many interacting ``agents'' (virtual entities representing in a
schematic way human beings, animals or bacteria for instance)
\cite{Schelling,Toner,Castellano}.
In these cases, interaction rules do not derive from an energy
function as in physical systems, but still correspond to direct
interactions between agents.

There however exist interesting situations in which agents have no direct
interactions between them, but rather interact with an external system, having
its own dynamics, and producing a feedback on the agents.
A simple example of such a situation is the case of an assembly
of customers who go for shopping on a regular time basis --say, every
week-- and need to choose between several equivalent stores.
In the absence of bias between shops, that is if customers choose any of the shops
with equal probability, all stores should maintain the same activity
level. If however, a fluctuation of activity occurs, one shop
may become slightly less attractive because, for instance, the sold products
get older (which may have significant consequences when selling, e.g.,
fresh food).
One can intuitively expect that this feedback loop may, in some cases,
result in an amplification mechanism leading to a symmetry breaking between shops.

Modelling the competition between stores is a difficult issue, which raised
interest in the economic literature from different perspectives
\cite{Ryan,Krider,Bhatnagar,Janssen,Hotling}
as well as, marginally, in the physics literature \cite{Jensen}.
Such models are often grounded in discrete choice theory \cite{Palma,Mira}.
Despite the central role played by prices in economic modelling,
empirical studies show that pricing mechanisms for fresh products
like fruits and vegetables are complex, and that other criteria related
to the quality of the products also play
an important role in customers decisions \cite{McLaughlin}.
The freshness of products thus turns out to be an important ingredient to take
into account when modelling shops selling perishable goods
\cite{Krider,Bhatnagar,Tsiros}.

In the present paper, we investigate from a statistical physics viewpoint
a simple model in which the ageing of products,
and the corresponding decrease of utility, are the central ingredients.
At variance with economic models, we neglect all financial aspects
(like the price of products and the fixed costs that the shopkeeper has to pay),
in order to single out a unique dynamical feedback mechanism
(the freshness of products)
between shops and customers, thus leading to a minimal dynamical model of shop
competition and decision making. Focusing on the symmetry breaking phenomenon,
we find both through numerical simulations and analytical mean-field approaches
that a phase transition takes place between a symmetric state in which the
two stores maintain the same level of activity, and a symmetry broken state
where a majority of customers choose a given store, in which they can find
fresher products.

\section{Model of competing stores}

\subsection{Definition}

The model is defined as follows. A set of $N_a$ agents randomly visit, on a regularly
time basis, one of two competing stores selling the same products.
Each store has the same number $N_p$ of products, which are all identical
except for their age. The age $\tau$ defines the freshness $h(\tau)$ of the corresponding
product, where $h(\tau)$ is a decreasing function, taking values between $0$ and $1$,
of the variable $\tau \ge 0$. A typical functional form for $h(\tau)$ is
\be \label{exp-aging}
h(\tau) = e^{-\tau/\tau_1}
\ee
where $\tau_1$ is the characteristic ageing time of products.

At discrete time steps $t_k=k \delta t$, with $\delta t=\tau_0/N_a$, an agent is chosen
at random with uniform probability. This agent $i$ then randomly selects
one of the two stores with a probability related to its satisfaction $S_{ij}$
with respect to each store $j=1,2$.
To be more specific, the probability $p_i$ that agent $i$ chooses the first store
is given by
\be \label{def-pi}
p_i = \frac{1}{1+e^{-\Delta S_i/T}}
\ee
where $\Delta S_i=S_{i1}-S_{i2}$ is the satisfaction difference of agent $i$
with respect to the two shops,
and $T$ is a parameter playing the role of a temperature, taking into account
the influence of other factors that are not described explicitly.
The second store is chosen with probability $1-p_i$. This dynamical rule preserves the symmetry between the two shops.
Note that Eq.~(\ref{def-pi}) is known as the logit rule in the socio-economic
literature \cite{Palma}, and that it is also formally similar to the Glauber transition
probability used in physics, provided $-S_i$ can be interpreted as an energy
(which is not necessarily the case in social models, see e.g.~\cite{pnas}).

Once agent $i$ has chosen a store $j=1,2$, it visits this store and buys
one product chosen at random (we recall that only one type of product is sold,
only the age $\tau$ differs from one to the other).
The sold product is immediately replaced by a fresh one, with age $\tau'=0$.
The agent a posteriori assesses the freshness $h(\tau)$ of the bought product, and
updates its satisfaction with respect to the store $j$ according to
\be \label{sat-update}
S_{ij}' = \alpha S_{ij}+(1-\alpha) h(\tau),
\ee
where $0\le\alpha<1$ is a parameter characterizing the agents' memory (all agents
have the same value of $\alpha$).
Then the process is iterated and a new agent is chosen randomly.

\begin{figure}[t]
\centering\includegraphics[width=15cm,clip]{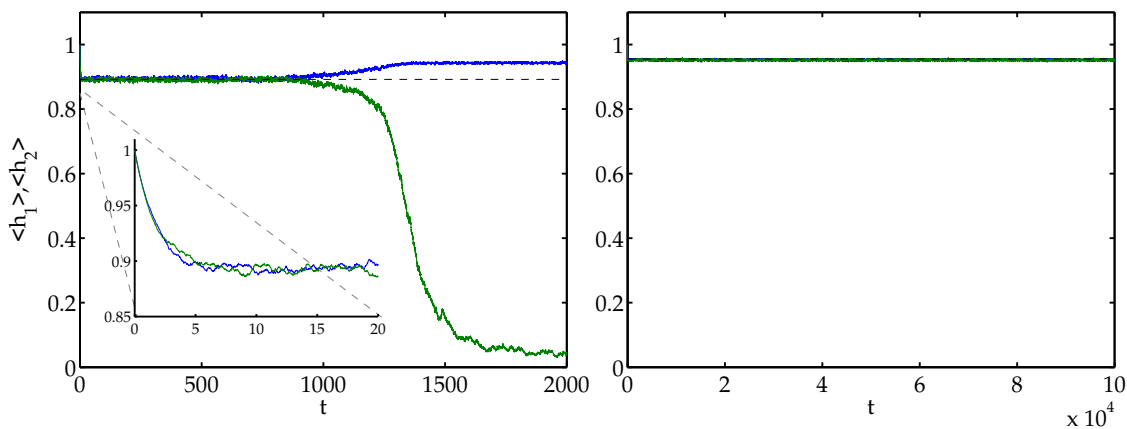}
\caption{Illustration of the dynamical behaviour of the model, in cases
when symmetry breaking occurs (left) or does not occur (right);
the mean freshness $\langle h_j \rangle$ of products in each shop $j=1,2$
is plotted as a function of time. The inset on the left plot shows a short
time-window corresponding to the transient state of the simulation starting
from a non-equilibrated initial conditions.
Left panel: $\tau_1=16.7$; right panel: $\tau_1=40$.
The other parameters are the same on both panels: $N_a=N_p=500$, $T=0.1$, $\tau_0=1$.}
\label{fig-ex}
\end{figure}

\subsection{Qualitative behaviour of the model}

Simulating numerically the above model, we observe for low enough temperature
a symmetry breaking between the two stores
(see the left panel of Fig.~\ref{fig-ex}): one of the stores
reaches a high level of commercial activity, resulting in a high freshness of the sold
products, while the second one only maintains a low activity and a low freshness
level. Interestingly, symmetry can be restored by varying the parameter $\tau_1$,
which controls the ageing rate of products, as seen on the right panel of
Fig.~\ref{fig-ex}; both shops then keep the same level of freshness.
Less surprisingly, symmetry is also restored at high temperature (not shown).

In the following sections, we quantitatively study this symmetry breaking
phase transition, through both analytical and numerical methods.
Section~\ref{sec-mean-field} is devoted to a mean-field approach
that allows us to obtain the phase diagram of the model. We find 
in particular a transition line in the plane defined by the temperature
$T$ and a control parameter $R$ related to the characteristic ageing
time $\tau_1$. Section~\ref{sec-simus} is then devoted to the comparison
between mean-field results and numerical simulations.
Finally, Section~\ref{sec-discuss} briefly discusses an analogy with
the Ising model, one of the most standard model exhibiting symmetry breaking.

\section{Mean-field approach}
\label{sec-mean-field}

\subsection{Framework and basic dynamical equations}

In this section, we present an analytical approach of the above model,
in order to characterize quantitatively the phase transition between
symmetric and non-symmetric states of the two stores.
Our analytical framework relies on
the simplifying assumption that the fluctuations of the agents' satisfaction
can be neglected, so that all agents are considered to have
a satisfaction difference $\Delta S$
equal to the average value $\overline{\Delta S}$.
In socio-economic terms, one would say
that the set of agents is replaced by a ``representative agent'' \cite{Palma}.
To lighten the notations, we introduce $\sigma \equiv \overline{\Delta S}$.
From Eq.~(\ref{sat-update}), we get, assuming that the time spent between two
visits to a shop is exactly $\tau_0$ (while this is only true on average):
\be
\sigma(t+\tau_0) = \alpha \sigma(t) + (1-\alpha) \left(\overline{h}_1(t)
- \overline{h}_2(t) \right),
\ee
where $\overline{h}_j(t)$ is the average freshness of products in the store $j$
at time $t$.
This equation can be reformulated as
\be \label{eq-sigma1}
\frac{1}{\tau_0}\, \big( \sigma(t+\tau_0) - \sigma(t) \big) = \frac{1-\alpha}{\tau_0}
\left( -\sigma(t) + \overline{h}_1(t) - \overline{h}_2(t) \right).
\ee
Hence $\sigma(t)$ varies significantly over a time scale of the order of
$\tau_0/(1-\alpha)$. In the limit of strong memory, $0<1-\alpha \ll 1$,
$\tau_0/(1-\alpha) \gg \tau_0$ so that the variations of $\sigma$ are very small
over a time $\tau_0$. Hence $\sigma(t)$ can be considered as a continuous function
of the variable $t$, and the l.h.s.~of Eq.~(\ref{eq-sigma1}) can be replaced
by a time derivative. We thus end up with
\be \label{eq-sigma2}
\frac{d\sigma}{dt} = \gamma \left( -\sigma(t) + \overline{h}_1(t)-\overline{h}_2(t) \right),
\ee
with $\gamma=(1-\alpha)/\tau_0$.
It is convenient to rather express the dynamics in terms of the fraction $p$ of
customers choosing the first store, instead of keeping the variable $\sigma$.
Using Eq.~(\ref{def-pi}) and replacing $\Delta S_i$ by the average value $\sigma$,
we get the relation
\be \label{eq-sigma-p}
\sigma = -T \ln \left(\frac{1-p}{p}\right),
\ee
which can be plugged into Eq.~(\ref{eq-sigma2}), leading to
\be \label{eq-dpdt1}
\frac{dp}{dt} = \frac{\gamma}{T}\, p(1-p) \left[ T \ln \left(\frac{1-p}{p}\right)
+ \overline{h}_1(t)-\overline{h}_2(t) \right].
\ee
We now need to compute the average freshness $\overline{h}_j$ as a function of
$p$. To this aim, we first determine the age distribution of products.

\subsection{Age distribution of products}

Let us denote as $\lambda_j(t)\, dt$ the average fraction of products sold
in store $j$ in the time interval $[t,t+dt]$,
and $\phi_j(\tau,t)$ the probability distribution of the age $\tau$
of products in store $j$ at time $t$, normalized
according to $\int_0^\infty \phi_j(\tau,t)\,d\tau=1$.
This distribution evolves in time according to the following relations:

\be
\label{evol-phi}
\cases{\phi_j(\tau+dt,t+dt) = \phi_j(\tau,t) - \lambda_j(t)\, dt\, \phi_j(\tau,t)& for $\tau > 0$,\\
\phi_j(0,t)\, dt = \lambda_j(t)\, dt.&\\}
\ee
The second equation in (\ref{evol-phi}) encodes the fact that sold products,
selected at random, are immediately replaced by new ones with age $\tau=0$.
Expanding the first equation in (\ref{evol-phi}) to linear order in $dt$,
one finds

\be \label{eq-fj2}
\frac{\partial \phi_j}{\partial \tau}(\tau,t) + \frac{\partial \phi_j}{\partial t}(\tau,t)
= -\lambda_j(t)\, \phi_j(\tau,t).
\ee
Note that integrating this equation over $\tau>0$ yields
$\phi_j(0,t)=\lambda_j(t)$, consistently with the second equation
in (\ref{evol-phi}).
The stationary solution of Eq.~(\ref{eq-fj2}) is exponential, namely
\be
\phi_j^{\mathrm{st}}(\tau) = \lambda_j\, e^{-\lambda_j \tau}.
\ee
Using the exponential ageing function given in Eq.~(\ref{exp-aging}),
the corresponding average freshness is thus
\be
\overline{h}_j = \int_0^{\infty} d\tau \phi_j^{\mathrm{st}}(\tau)\, e^{-\tau/\tau_1}
= \left( 1+\frac{1}{\lambda_j \tau_1} \right)^{-1}.
\ee
In addition, the fraction $\lambda_j \, dt$ of products sold in the time
interval $[t,t+dt]$ is obtained as the ratio of the number $N_a\, p_j\, dt/\tau_0$
of customers visiting shop $j$ in this interval, and of the number $N_p$
of products, yielding
\be
\lambda_j = \frac{N_a p_j}{N_p \tau_0}
\ee
with $p_1=p$ and $p_2=1-p$.
We thus end up, as far as the average freshness is concerned, with
\be
\overline{h}_j = \left( 1+\frac{R}{2p_j} \right)^{-1}
\ee
where we have defined the control parameter
\be
R = \frac{2N_p\tau_0}{N_a\tau_1}\;.
\ee
The dimensionless ratio $R$ compares $N_p$,
the total number of products available in a shop,
to $\frac{N_a}{2\tau_0}\, \tau_1$, the typical number of agents visiting
a shop during the characteristic ageing time of the products
(assuming a symmetric state).
A small value of $R$ thus corresponds to a situation where products are often renewed, while for large $R$ products are rarely selected and become old on average.
In the following, we use as an approximation the above steady-state distribution
of the age of products, assuming that the satisfaction of agents evolves
on time scales much longer than the turn-over time of products.
This assumption has, however, no consequence on the determination of the
fixed points of the dynamics.

\begin{figure}[t]
\centering\includegraphics[width=8cm,clip]{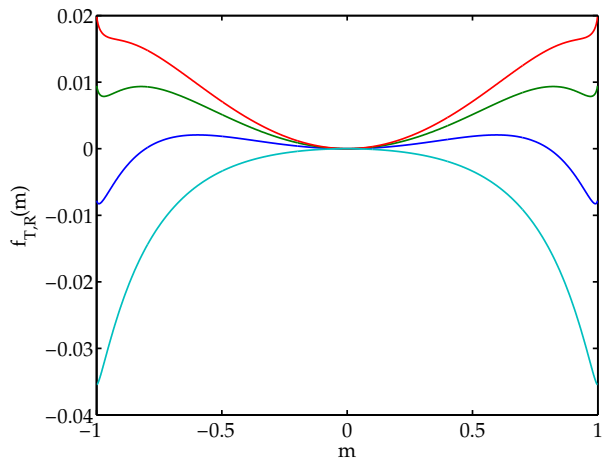}
\caption{Illustration of the behaviour of the free energy $f_{T,R}(m)$,
for four different temperatures (from top to bottom: $T=0.2$, $0.185$,
$0.16$, and $0.12$) and the same value $R=0.2$, showing the onset of the two
symmetric secondary minima.}
\label{fig-free-energy}
\end{figure}

\subsection{Determination of the phase transition}
\label{subsec-phase-trans}

Gathering results, we find for the time evolution of $p$ the following equation

\be \label{eq-dpdt2}
\frac{dp}{dt} = \frac{\gamma}{T}\, p(1-p) \left[ T \ln \left(\frac{1-p}{p}\right)
+ \left( 1+\frac{R}{2p} \right)^{-1} - \left( 1+\frac{R}{2(1-p)} \right)^{-1} \right].
\ee
Introducing an order parameter $m$ through $p=(1+m)/2$,
we can formally rewrite Eq.~(\ref{eq-dpdt2}) as
\be \label{dmdt}
\frac{dm}{dt} = \frac{\gamma}{T}\, (1-m^2)\, g_{T,R}(m),
\ee
with $g_{T,R}(m)$ given by
\be
g_{T,R}(m) = \frac{1}{2} \left[ T \ln \left(\frac{1-m}{1+m}\right) +
\left( 1+\frac{R}{1+m} \right)^{-1} - \left( 1+\frac{R}{1-m} \right)^{-1} \right].
\ee
The reason for excluding the factor $\gamma (1-m^2)/T$ from the definition
of $g_{T,R}(m)$ will be discussed in Sec.~\ref{sec-discuss}.
In order to study possible phase transitions, it is convenient to
define an effective free energy (or potential function) $f_{T,R}(m)$ through
\be
f_{T,R}(m) = -\int_0^m g_{T,R}(m') \, dm',
\ee
so that
\be \label{dmdt2}
\frac{dm}{dt} = -\frac{df_{T,R}}{dm}(m).
\ee

\begin{figure}[t]
\centering\includegraphics[width=9cm,clip]{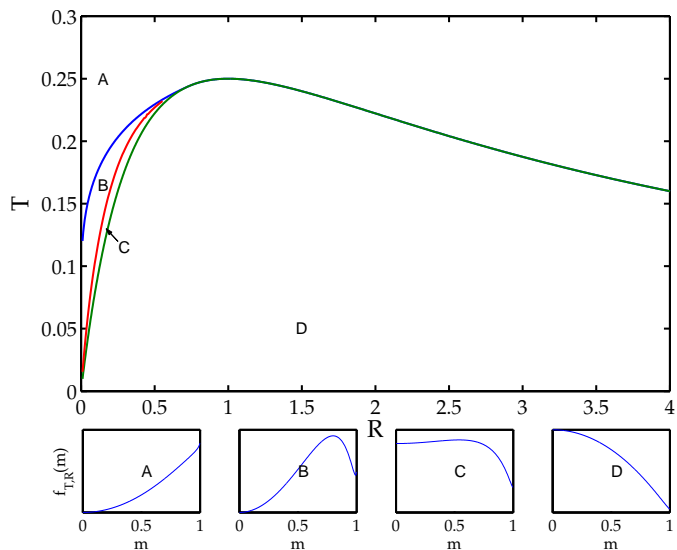}
\caption{Mean-field phase diagram, with four different regions. A: $m=0$ is the only minimum of $f_{T,R}(m)$. B: $m=0$ is the global minimum, but secondary minima have appeared. C: $m=0$ is only a local minimum, the two symmetric minima are the most stable. D: $m=0$ is a local maximum, the two symmetric minima are the only stable states.
The transition lines between those regions are respectively defined by $T(R)=T_s$
between A and B, $T_d$ between B and C, and $T_c$ between C and D. $T_c$ also
separates A and D for $R>R^* \approx 0.732$, when the transition is continuous.
The lower panels show the typical shape of $f_{T,R}(m)$
for $m>0$ ($f_{T,R}(m)$ is a symmetric function) in each region.}
\label{fig-diag-theo}
\end{figure}

\noindent
It is easily seen that $m=0$, corresponding to the symmetric state,
is a steady-state solution of Eq.~(\ref{dmdt}).
To test the local stability of this solution,
we expand the free energy $f_{T,R}(m)$ for small $m$, to order $m^2$,
leading to
\be
f_{T,R}(m) = \frac{1}{2} \left[ T-\frac{R}{(1+R)^2} \right] m^2
+ \mathcal{O}(m^4).
\ee
The solution $m=0$ is linearly stable if the bracket on the r.h.s.~is
positive. The critical temperature $T_c$ is reached when the bracket vanishes,
yielding
\be
T_c = \frac{R}{(1+R)^2}.
\ee
When this transition coincides with the onset of two symmetric minima for
$T<T_c$, the transition is continuous.
For low values of $R$, however, two symmetric secondary minima appear
already for a value $T_s > T_c$, when $m=0$ is still the global minimum
as illustrated on Fig.~\ref{fig-free-energy}.
This case corresponds to a discontinuous transition, which should be
reached (in analogy to the equilibrium theory of phase transitions)
at a temperature $T_d$ such that all three minima have equal values
$f_{T,R}(m)=0$.

The resulting mean-field phase diagram is presented
on Fig.~\ref{fig-diag-theo}. In region A, a single steady-state solution ($m=0$)
exists, while in region D, only the two symmetric solutions remain.
In region B, between the lines $T_s$ and $T_d$, the symmetric
solution $m=0$ is the most stable, while in region C, between $T_d$ and $T_c$,
the most stable solutions are the non-zero ones.
The lines $T_s$, $T_d$ and $T_c$ meet at a value $R^*=\sqrt{3}-1 \approx 0.732$.

The nature of the discontinuous transition however calls for a short comment.
First of all, we note that the evolution
equation (\ref{dmdt}) is deterministic, so that the system should remain
trapped forever once it has reached a locally stable stationary
solution. However, in a more realistic description, the finite size
of the system should result in a small effective noise, making the
system explore the vicinity of the stationary state. In this case,
the most stable stationary states correspond to the absolute minima
of $g_{T,R}(m)$ (either $m=0$, or two symmetric values $m_0$ and $-m_0$).

\begin{figure}[t]
\centering\includegraphics[width=11cm,clip]{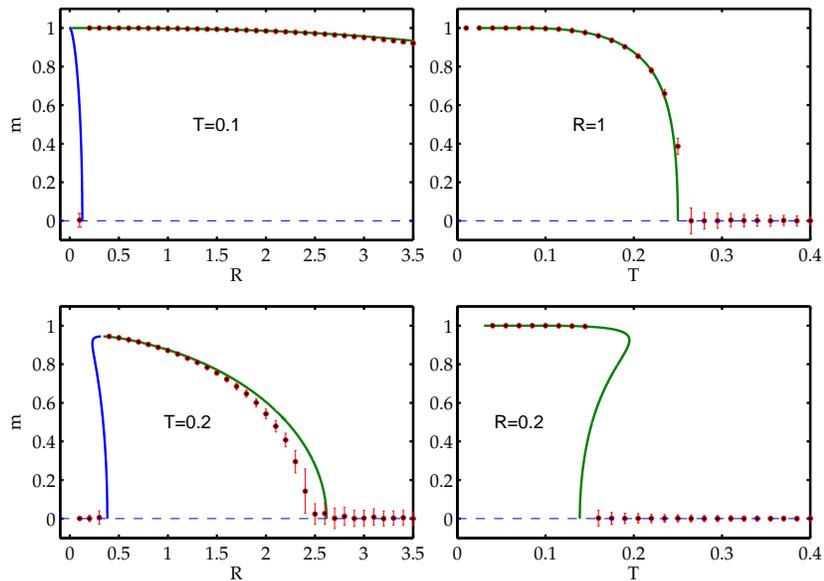}
\caption{Order parameter $m$, plotted either as a function
of $R$ or as a function of temperature $T$.
Lines correspond to the mean-field predictions,
while dots with error bars are the results from numerical simulations.
On all four figures, a phase transition, either continuous or discontinuous,
is observed between a symmetric phase (both stores have the same activity)
and a phase with broken symmetry (one of the stores has a higher activity level).}
\label{fig-theo-num}
\end{figure}

\section{Comparison between mean-field and numerical results}
\label{sec-simus}

\subsection{Measure of the order parameter}

We now turn to the results of numerical simulations of the agent-based model,
and compare them with mean-field results. The numerical value for $m$ is obtained as a time average (over $10^4 \,\tau_0$) for a simulation usually involving $N_a=500$ agents and as many products as agents ($N_p=N_a$).
The memory parameter is set to $\alpha=0.8$, a value sufficiently large
for the continuous time description used in Eq.~(\ref{eq-sigma2})
to be approximately valid.
We have however checked that setting this parameter in the range
$0.5 \le \alpha \le 0.95$ only changes the numerical value of $m$
by less than a few percent for most values of $R$ and $T$.

We first plot on Fig.~\ref{fig-theo-num} the steady-state value of
the order parameter $m$
as a function of temperature $T$ and parameter $R$.
An overall good agreement is observed with the mean-field predictions,
that are obtained as follows.
The fixed points of Eq.~(\ref{dmdt}) are solutions of $g_{T,R}(m)=0$.
Besides the trivial solution $m=0$, a solution $m_0>0$ exists 
below the transition line. An analytical expression for $m_0(T,R)$ is
hard to obtain, but one can instead easily express $T$ as a function
of $m$ and $R$:
\be
T(m,R) = \frac{ \left( 1+\frac{R}{1+m} \right)^{-1}-\left( 1+\frac{R}{1-m} \right)^{-1}}{\ln(1+m)-\ln(1-m)}
\ee
Fixing $R$, it is then straightforward to plot $T$ as a function of $m$.
One can also invert $T(R,m)$ to get $R(T,m)$, as shown on Fig.~\ref{fig-theo-num}.

\subsection{Determination of the transition line}

\begin{figure}[t]
\centering\includegraphics[width=15cm,clip]{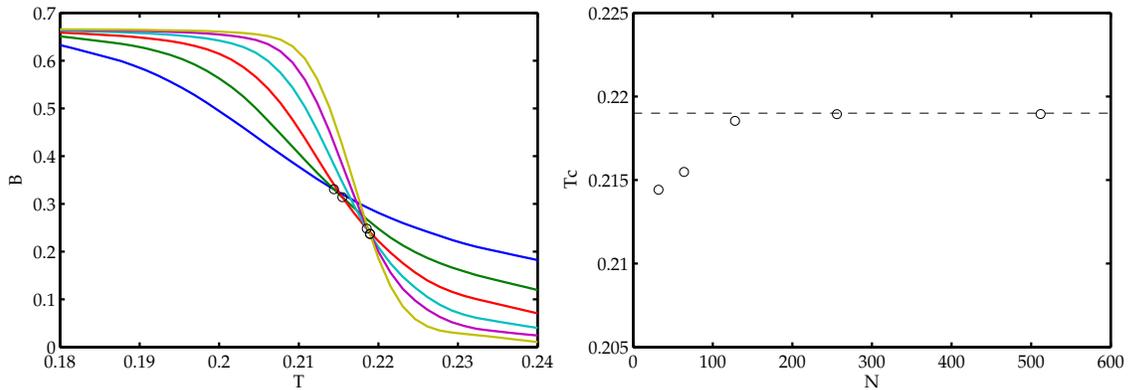}
\caption{Left: behaviour of the Binder cumulant $B$
as a function of temperature $T$ for different sizes
$N_a=16$, $32$, $64$, $128$, $256$, $512$ ($R=2$, simulation time:
$5\times 10^5 \,\tau_0$).
The critical point is obtained as the large $N$ limit of the
intersections of successive curves.
Right: crossing point of two successive curves $N_a=N/2$ and $N_a=N$,
as a function of $N$.}
\label{fig-binder}
\end{figure}

\begin{figure}[t]
\centering\includegraphics[width=10cm,clip]{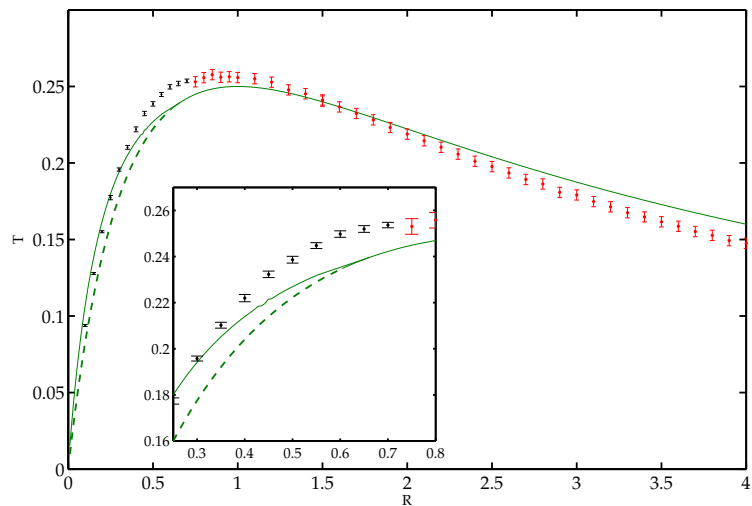}
\caption{Phase diagram (T-R), showing the transition line
between the symmetric phase and the symmetry-broken phase.
Lines are mean-field predictions, and dots correspond to the phase transition
observed numerically, from the order parameter curve
for $R<R^* \approx 0.732$ (discontinuous transition) and
from the Binder cumulant for $R>R^*$ (continuous transition).
The mean-field transition line (either continuous or discontinuous)
is indicated by a full line; it corresponds to $T_c$ for
$R>R^*$ and to $T_d$ (set of points for which the three minima
of the free energy have the same value) for $R<R^*$.
In the region $R<R^*$, $T_c$ is indicated as a dashed line.
The inset shows a zoom on the region where $T_c$ and $T_d$ intersect.}
\label{fig-phase-diag}
\end{figure}

We also compare on Fig.~\ref{fig-phase-diag} numerical results for
the phase diagram with the mean-field predictions.
To determine the transition line, we have distinguished two methods for the
cases of continuous and discontinuous phase transitions.
For the discontinuous case ($R<R^*$), the transition point can be
determined from the jump of the order parameter $m$.
In the continuous case, the transition point
is more difficult to determine precisely from the average value of the
order parameter, due to finite size effects.
We thus have used the standard technique of
the Binder cumulant, which characterizes the deviation from
Gaussianity of the probability distribution of the order parameter
\cite{Binder}. For a centered variable $m$, the Binder cumulant
is defined as
\be
B = 1- \frac{\langle m^4 \rangle}{3\langle m^2 \rangle^2}.
\ee
The main interest of the Binder cumulant is that when plotted
as a function of the control parameter, curves corresponding
to different system sizes intersect at the critical point,
with only small finite size effects.
This method allows for a rather precise estimate of the transition point.
An illustration of the behaviour of the Binder cumulant is given
on Fig.~\ref{fig-binder}.

Altogether, a rather good agreement is obtained
between mean-field and numerical results, though some systematic
deviations appear, as seen on Fig.~\ref{fig-phase-diag}.
These deviations might be due to underevaluated finite size effects
(the number of agents used in the simulation is $N_a \sim 500$),
but it is more likely to result from the simplifying assumptions
made in the derivation the mean-field equations.
Note however that the ranges in which continuous and discontinuous transitions
are observed numerically are rather well reproduced by the mean-field argument.

\section{Discussion}
\label{sec-discuss}

In this section, we briefly discuss the relation between the present
model and one of the most standard models exhibiting
a symmetry breaking phase transition, namely the Ising model, before
drawing some conclusions and perspectives.

\subsection{Relation to the Ising model}

In the present model, each agent has to choose between two options,
namely going to the first or the second store.
Such a binary choice can be encoded into a 'spin' variable $s_i$,
equal to $1$ if agent $i$ chooses the first shop, and to $-1$
in the opposite case. Though somehow artificial, this definition
has the advantage that the order parameter $m$ introduced in
Sec.~\ref{subsec-phase-trans} can be reexpressed as
$m=N_a^{-1} \sum_i s_i$.
This naturally suggests that the present model may exhibit, to some
extend, an analogy to the Ising model.
This analogy can be made more precise, at the mean-field level,
in the following way. Let us modify the model in such a way
that the freshness of products (which drives the evolution of
the satisfaction of agents) is replaced by the opposite of the energy
$E_{\mathrm{Ising}}$ of the mean-field Ising model,
with $E_{\mathrm{Ising}}=-\frac{1}{2}N_a J m^2$
\footnote{Strictly speaking, the magnetization $m$ should be computed
from the last choice of the $N_a-1$ other agents.}.
Then the mean satisfaction $\sigma$ introduced
in Sec.~\ref{sec-mean-field} evolves according to
\be
\sigma(t+\tau_0) = \alpha \sigma(t) -(1-\alpha) \Delta E,
\ee
with
\be
\Delta E = -\frac{1}{2} N_a J \left[ \left(m+\frac{1}{N_a}\right)^2
- \left(m-\frac{1}{N_a}\right)^2 \right] = -2Jm.
\ee
Note that the last equality has been obtained in the large $N_a$ limit.
We then get the evolution equation, in the large memory limit
$1-\alpha \ll 1$,
\be
\frac{d\sigma}{dt} = \gamma [-\sigma + 2Jm],
\ee
which is the analog of Eq.~(\ref{eq-sigma2}), with here again
$\gamma=(1-\alpha)/\tau_0$.
It is easy to check, using relation (\ref{eq-sigma-p}) to eliminate
the variable $\sigma$, that the time evolution equation for $m$
is given by
\be
\frac{dm}{dt} = \frac{\gamma}{T} (1-m^2) \left[ \frac{T}{2}\ln\left(\frac{1-m}{1+m}\right)+Jm \right].
\ee
Similarly to Eq.~(\ref{dmdt}), the term between brackets can be written
as the opposite of the derivative of a free energy $f(m)$,
\be
f(m) = \frac{T}{2}(1+m)\ln(1+m) + \frac{T}{2}(1-m)\ln(1-m)
- \frac{1}{2} Jm^2
\ee
which is, up to an additive constant $T\ln 2$, nothing but the standard mean-field free energy of the Ising model.

However, beyond mean-field, it is clear that the relation to the Ising model
is at most an analogy. Indeed, the fact that the interaction between agents
is here mediated by the ageing of products in each store rather than
by an internal energy function makes the present model quite different from
standard equilibrium models.

\subsection{Conclusion and perspectives}

In this paper, we have studied a simple model of competing shops
selling fresh products, retaining as main ingredients the ageing of products
and the satisfaction of customers with respect to the freshness of products
--considered to be the main factor influencing choices made by customers.
Under these very simplifying assumptions, we found a surprisingly rich
dynamics, with both continuous and discontinuous transitions between
symmetric and symmetry-broken states. The overall agreement between
numerical simulations and mean-field predictions was found to be
satisfactory.

Several extensions could be interesting to study.
For instance, other functional forms could be considered for
the ageing function $h(\tau)$ introduced in Eq.~(\ref{exp-aging}), like
\be \label{lin-aging}
h(\tau) = \left(1-\frac{\tau}{\tau_1}\right)\, \theta \left(1-\frac{\tau}{\tau_1}\right)
\ee
where $\theta(x)$ is the Heaviside function. Such a form would be rather
natural for products that can be sold only up to an expiration date.
We have checked that the form (\ref{lin-aging})
preserves the qualitative behaviour of the model.
Another natural extension of this work would be to include more than two shops.
Although we did not study this case in detail, a simple argument
can already be given here.
With a fixed overall number of agents $N_a$, adding
shops reduces the nominal number of customers for a given shop.
As a first approximation, a model with more than two shops can be described
as a two-shop model with an effective number of agents $N_a^{\mathrm{eff}} < N_a$,
resulting in an effective parameter $R^{\mathrm{eff}}>R$.
If one starts from a low value of $R$ corresponding to
the symmetric region, increasing the number of shops would then eventually
lead to a value $R^{\mathrm{eff}}$ in the symmetry broken phase. It is thus likely
that, for a given total number of agents, there is a maximal number of shops
that can maintain a good level of activity, which seems to be
consistent with conventional wisdom.

Further possible extensions may include taking into account
explicitly the prices of products, as well as the costs associated
to the shops (e.g., rent or salaries).
In this context, one can imagine to include, in a simplified way,
a dynamics of the shopkeeper who could modify the prices, or diversify
the products, as a function of the dynamical state of the store.
One might guess that such a dynamics could contribute to stabilize the
symmetric states.

\end{document}